# Landau damping

John Wesson

*CCFE, Culham Science Centre,
Abingdon, Oxfordshire OX14 3DB, UK*

**Abstract.** Landau damping is calculated using real variables, clarifying the physical mechanism.

Landau's calculation of electron plasma oscillations demonstrated the phenomenon now known as Landau damping[1]. The calculation used a Fourier-Laplace transform and regarded the electron velocity as complex in order to properly locate the pole in the complex frequency plane that gives the dispersion relation containing the wave damping.

Here, the same subject is examined using real variables and straightforward algebra. The aim is to understand the damping and growth of plasma oscillations, shown by Landau to arise from particles close to resonance with the plasma wave. It is found that the physics underlying the damped oscillations is different from that in the case of growth.

The governing equations for both growth and damping are the linearised Vlasov equation for the distribution function $f_1$

$$\frac{\partial f_1}{\partial t} + v\frac{\partial f_1}{\partial x} = \frac{E_1 e}{m}\frac{\partial f_0}{\partial v} \qquad (1)$$

where $E_1$ is the electric field, and Poisson's equation

$$\frac{\partial E_1}{dx} = -\frac{e}{\varepsilon_0}n_1 = -\frac{e}{\varepsilon_0}\int f_1 dv. \qquad (2)$$

For simplicity the wavelength will be taken to be much larger than the Debye length, so that $kv_T \ll \omega_p$ where k is the wave number, $v_T$ the electron thermal velocity and $\omega_p$ the plasma frequency.

Since the Landau effect arises from particles travelling close to the wave velocity, the algebra is more transparent if the calculation is carried out in the frame of the wave, as illustrated in Fig.1. Damping and growth depend on the sign of $(\partial f_0/\partial v)$ at the wave velocity and as the behaviour is different for the cases $(\partial f_0/\partial v)_w > 0$ and $(\partial f_0/\partial v)_w < 0$, the calculations will be made clearer by treating them separately.

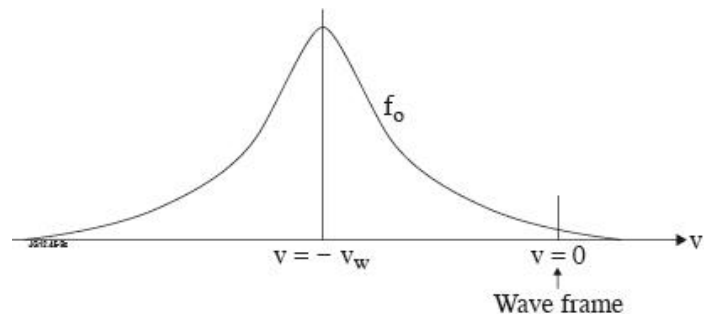

Fig.1. Illustrating the choice of frame. $v_w$ is the wave velocity in the frame of the plasma.

## The case with $(\partial f_0/\partial v)_w > 0$

With $(\partial f_0/\partial v)_w > 0$ there are homogeneous solutions for which all of the terms in eqns 1 and 2 have a factor $e^{\gamma t}$.

So the electric field can be written

$$E_1 = \hat{E} \sin kx \, e^{\gamma t} \qquad (3)$$

and the distribution function takes the form

$$f_1 = \left(\hat{f}_s \sin kx + \hat{f}_c \cos kx\right) e^{\gamma t}. \qquad (4)$$

Substituting eqns 3 and 4 into eqn 1 and equating sine terms and cosine terms leads to the solution

$$f_1 = \frac{\hat{E}e}{m}\left(\frac{\gamma}{\gamma^2 + k^2 v^2}\frac{\partial f_0}{\partial v}\sin kx - \frac{kv}{\gamma^2 + k^2 v^2}\frac{\partial f_0}{\partial v}\cos kx\right) e^{\gamma t}. \qquad (5)$$

In calculating $n_1$ there are two contributions from $f_1$. One is the contribution, $n_1^w$, localised around $v = 0$, coming from velocities for which $v \sim \gamma/k$. The other, $n_1^b$, comes from the basic thermal distribution around $v = -v_w$. Since $v_T \ll v_w$ and anticipating $\gamma/k \ll v_w$, the two contributions are well separated and can be treated independently. The contribution $n_1^b$ is obtained by taking $\gamma \ll kv_w$, writing

$$n_1^b = \frac{\hat{E}e}{m}\left(\int \frac{\gamma}{k^2v^2}\frac{\partial f_0}{\partial v}dv\sin kx - \int \frac{1}{kv}\frac{\partial f_0}{\partial v}dv\cos kx\right)e^{\gamma t}$$

and integrating by parts to obtain

$$n_1^b = \frac{\hat{E}e}{m}\left(\int \frac{2\gamma}{k^2v^3}f_0 dv\sin kx - \int \frac{1}{kv^2}f_0 dv\cos kx\right)e^{\gamma t}.$$

The leading order terms in the required integrals are obtained by taking $f_0$ to have the form of a delta function at the velocity $-v_w$ in the wave frame. Thus

$$n_1^b = \frac{\hat{E}en}{m}\left(-\frac{2\gamma}{k^2v_w^3}\sin kx - \frac{1}{kv_w^2}\cos kx\right)e^{\gamma t}. \qquad (6)$$

In the localised contribution from particles with velocities close to the wave velocity, $\partial f_0/\partial v$ can be taken to be constant, and the resulting contribution to $n_1$ is

$$n_1^w = \frac{\hat{E}e}{m}\left(\frac{\partial f_0}{\partial v}\right)_w \left(\int \frac{\gamma}{\gamma^2+k^2v^2}dv\sin kx - \int \frac{kv}{\gamma^2+k^2v^2}dv\cos kx\right)e^{\gamma t}$$

$$= \frac{\hat{E}e}{m}\left(\frac{\partial f_0}{\partial v}\right)_w \left(\frac{1}{k}\tan^{-1}\frac{kv}{\gamma}\bigg|_{kv\ll-|\gamma|}^{kv\gg|\gamma|}\sin kx + 0\right)e^{\gamma t}$$

$$= \frac{\hat{E}e}{m}\left(\frac{\partial f_0}{\partial v}\right)_w \frac{\gamma}{|\gamma|}\frac{\pi}{k}\sin kx\, e^{\gamma t}.$$

The integrand $\frac{\gamma}{\gamma^2+k^2v^2}$ has the form shown in Fig.2, the contribution to $n_1^w$ being independent of the magnitude of $\gamma$.

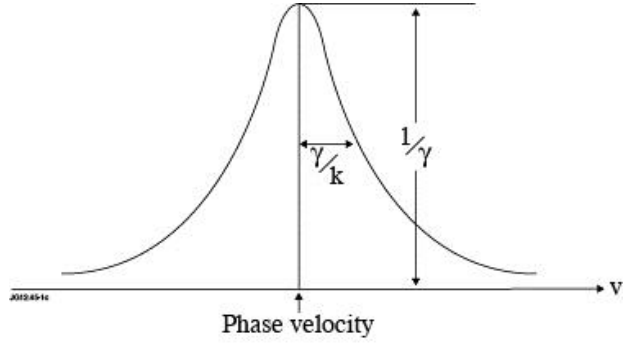

Fig. 2. The form of the velocity dependence of the density contribution around the phase velocity.

The complete $n_1$ is now given by $n_1^b + n_1^w$,

$$n_1 = \frac{e\hat{E}}{m}\left(\left(-\frac{2\gamma n}{k^2 v_w^3} + \frac{\gamma}{|\gamma|}\frac{\pi}{k}\left(\frac{\partial f_0}{\partial v}\right)_w\right)\sin kx - \frac{n}{kv_w^2}\cos kx\right)e^{\gamma t}$$

and the dispersion relation is obtained by its substitution into eqn 2 to obtain

$$\frac{1}{\omega_p^2}\cos kx = \left(\left(+\frac{2\gamma}{k^3 v_w^3} - \frac{\gamma}{|\gamma|}\frac{1}{n}\frac{\pi}{k^2}\left(\frac{\partial f_0}{\partial v}\right)_w\right)\sin kx + \frac{1}{k^2 v_w^2}\cos kx\right) \quad (7)$$

where $\omega_p = \left(ne^2/\varepsilon_0 m\right)^{1/2}$.

Equating the cos kx terms in eqn 7 gives the wave velocity $v_w = \pm\ \omega_p/k$ and for the present case

$$v_w = \frac{\omega_p}{k}.$$

The sin kx term in eqn 7 is zero, and this condition determines $|\gamma|$

$$|\gamma| = \frac{\pi\omega_p^3}{2k^2 n}\left(\frac{\partial f_0}{\partial v}\right)_w. \quad (8)$$

It is clear that eqn 8 does not allow a solution for $(\partial f_0/\partial v)_w < 0$, and for $(\partial f_0/\partial v)_w > 0$ the required solution is

$$\gamma = \frac{\pi \omega_p^3}{2k^2 n}\left(\frac{\partial f_0}{\partial v}\right)_w.$$

It is seen from the equation for $n_1^w$ that the contribution that leads to growth comes from sojourning particles which, during a growth time, $1/\gamma$, travel a distance $v/\gamma$ that is less than $2\pi/k$ and so do not sample the whole wave.

The sojourning particles do not make a direct growth of charge in phase with the basic charge of the wave, but produce an out-of-phase charge as illustrated in Fig. 3. This out-of-phase charge is balanced out by the modified contribution from the main particle distribution, which is proportional to $\gamma$. The balancing of these two contributions determines $\gamma$, as given by eqn. 7. The growth of the charge in the wave actually arises simply from the divergence of the basic "flow" of the particles in the main distribution passing through the wave.

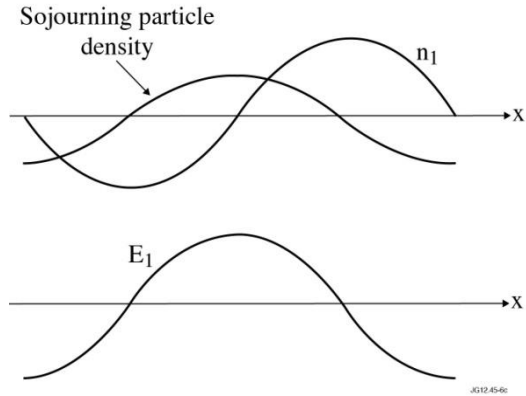

Fig.3. Sojourning particles produce an out-of-phase charge.

**The case with $(\partial f_0/\partial v)_w < 0$**

It is necessary now to determine the procedure that allows continuation of the case with $(\partial f_0/\partial v)_w > 0$ to that with $(\partial f_0/\partial v)_w < 0$.

It was seen from eqn 8 that homogeneous solutions are only possible with $(\partial f_0/\partial v)_w > 0$. For $(\partial f_0/\partial v)_w < 0$ the out-of-phase sin kx part of $n_1$ cannot be made zero and a solution with $\gamma < 0$ is not possible. This constraint arises from eqn 4 at the outset of the calculation.

Returning to eqn. 1, the treatment of $f_1$ in calculating the contributions to $n_1$ for $(\partial f_0/\partial v)_w < 0$ is unaffected except for the sin kx contribution for particles with velocities close to that of the wave. This contribution needs closer attention. With no assumed $e^{\gamma t}$ time dependence for $f_1$ it can be written

$$f_1 = f_s \sin kx + f_c \cos kx.$$

Substituting $f_1$ into eqn 1 leads to the equation for $f_s$

$$\frac{\partial^2 f_s}{\partial t^2} + k^2 v^2 f_s = \gamma \frac{e}{m} \hat{E} \frac{\partial f_0}{\partial v} e^{\gamma t}.$$

The calculation of $f_s^w$, the wave contribution to $f_s$, in the $(\partial f_0/\partial v)_w > 0$ case used the particular integral solution. In the case with $(\partial f_0/\partial v)_w < 0$ it is necessary to add the complementary function to obtain

$$f_s^w = \frac{\hat{E}e}{m} \left(\frac{\partial f_0}{\partial v}\right)_w \frac{\gamma}{\gamma^2 + k^2 v^2} e^{\gamma t} + \varphi(v) \cos(kvt).$$

The factor $\varphi(v)$ represents the initial form of the complementary function contribution to $f_s^w$. The procedure now is to determine the solution for $f_s^w$ that allows continuity with the $(\partial f_0/\partial v)_w > 0$ case. The first step is to choose $\varphi(v)$ to make the functional form of $f_s^w(v, t = 0)$ the same as that for $(\partial f_0/\partial v)_w > 0$. $f_s^w$ then becomes

$$f_s^w = \frac{\hat{E}e}{m} \left(\frac{\partial f_0}{\partial v}\right)_w \frac{\gamma}{\gamma^2 + k^2 v^2} \left(e^{\gamma t} + \alpha \cos(kvt)\right)$$

where $\alpha$ is a constant.

Noting that the contribution from the $\cos(kvt)$ term is localised around $v = 0$ and that, for $\gamma < 0$,

$$\int_{kv \ll -|\gamma|}^{kv \gg |\gamma|} \frac{\gamma}{\gamma^2 + k^2 v^2} \cos(kvt) \, dv = -\frac{\pi}{k} e^{\gamma t}$$

integration of $f_s^w(v)$ over v gives

$$n_1^w = -\frac{\hat{E}e}{m} \left(\frac{\partial f_0}{\partial v}\right)_w (1+\alpha) \frac{\pi}{k} \sin kx \, e^{\gamma t} \qquad \gamma < 0. \qquad (9)$$

Recalling that for $(\partial f_0/\partial v)_w > 0$

$$n_1^w = \frac{\hat{E}e}{m}\left(\frac{\partial f_o}{\partial v}\right)_w \frac{\pi}{k} \sin kx \, e^{\gamma t}, \qquad (10)$$

continuity of $n_1^w$ as expressed by eqns 9 and 10 requires $\alpha = -2$.

Adding the $n_1^b$ contribution, given in eqn 6, the complete sin kx component of $n_1$ now becomes

$$n_1^S = \frac{e\hat{E}}{m}\left(-\frac{2\gamma k n}{\omega_p^3} + \frac{\pi}{k}\left(\frac{\partial f_0}{\partial v}\right)_w\right)\sin kx \, e^{\gamma t} \qquad \gamma < 0$$

and putting this out-of-phase term to zero gives the damping rate for $(\partial f_0/\partial v)_w < 0$

$$\gamma = \frac{\pi \omega_p^3}{2k^2 n}\left(\frac{\partial f_0}{\partial v}\right)_w.$$

The mechanism of damping is not an inverse form of the sojourning particle mechanism responsible for growth. The additional part of the $f_1$ arising from the complementary function has introduced an out-of-phase component which has a time dependence cos(kvt). Initially cos(kvt) = 1, giving a full density contribution. As t increases the particles involved see increasingly different phases of the electric field and this phase mixing leads to damping, as shown in Fig. 4. Since the particles involved have v ~ $\gamma$/k, the characteristic time for phase mixing is 1/$\gamma$ as would be expected. There are equal damping contributions from those particles moving faster than the wave and those moving slower. The sojourning particles are still there but their contribution is outweighed by the damping due to phase mixing.

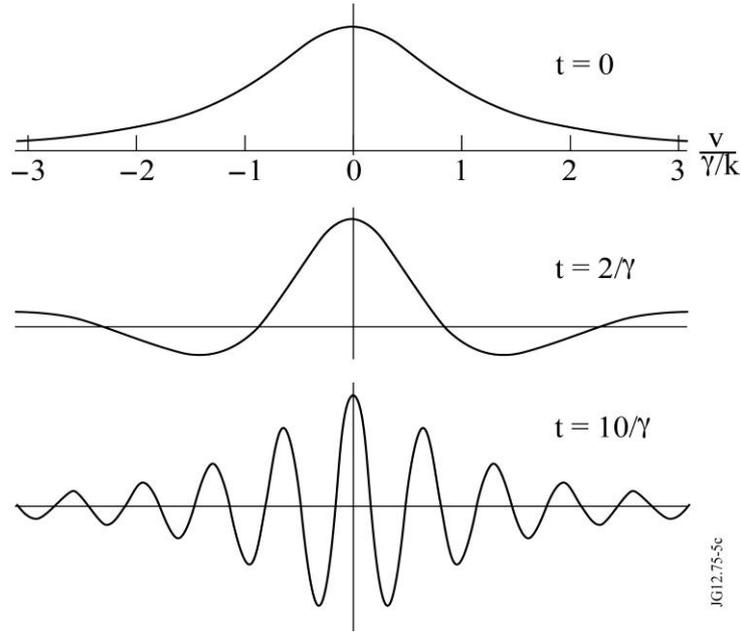

Fig. 4 The time development of the phase-mixing component of $f_1(v)$ is illustrated by its form at $t = 0$, $t = 2/\gamma$ and $t = 10/\gamma$.

**Overview**

Landau's treatment of plasma oscillations as an initial value problem, with an arbitrary $f_1(x,v,t = 0)$, leads to a solution for $E_1(x,t)$ that is a sum of modes. Given this solution it is then possible to calculate $f_1(k,v,t)$. The expression for $f_1(k,v,t)$ includes the arbitrary initial distribution function $f_1(k,v,t = 0)$.

In the present paper the aim is to analyse just the basic plasma oscillation, including the contribution of particles with velocities close to the wave velocity. This implies that there is a specific form of $f_1(v, t = 0)$ that gives the required solution. The cases with $(\partial f_0/\partial v)_w > 0$ and $(\partial f_0/\partial v)_w < 0$ require different initial conditions as shown below.

The initial form of $f_1$ can be written

$$f_1(v, t = 0) = f_1^s(t = 0) \sin kx + \frac{\hat{E}e}{m} \frac{\partial f_0}{\partial v} \frac{kv}{\gamma^2 + k^2 v^2} \cos kx$$

and the solutions $f_1(v, t)$ obtained in the preceding sections can then be written in the general form

$$f_1 = \left[ f_1^s(t=0)\cos(kvt) + \frac{\hat{E}e}{m}\frac{\partial f_0}{\partial v}\frac{\gamma}{\gamma^2+k^2v^2}\left(e^{\gamma t} - \cos(kvt)\right) \right] \sin kx$$
$$+ \frac{\hat{E}e}{m}\frac{\partial f_0}{\partial v}\frac{kv}{\gamma^2+k^2v^2} e^{\gamma t} \cos kx. \tag{11}$$

where, for $\left(\frac{\partial f_0}{\partial v}\right)_w \gtrless 0$,

$$f_1^s(t=0) = \pm\frac{\hat{E}e}{m}\frac{\partial f_0}{\partial v}\frac{\gamma}{\gamma^2+k^2v^2} \tag{12}$$

The specific form of the solutions obtained in previous sections is retrieved by substituting eqn 12 into eqn 11.

Thus, for $\left(\frac{\partial f_0}{\partial v}\right)_w > 0$

$$f_1 = \frac{\hat{E}e}{m}\frac{\partial f_0}{\partial v}\left(\frac{\gamma}{\gamma^2+k^2v^2}\sin kx + \frac{kv}{\gamma^2+k^2v^2}\cos kx\right)e^{\gamma t},$$

and for $\left(\frac{\partial f_0}{\partial v}\right)_w < 0$

$$f_1 = \frac{\hat{E}e}{m}\frac{\partial f_0}{\partial v}\left[\frac{\gamma}{\gamma^2+k^2v^2}\left(e^{\gamma t} - 2\cos(kvt)\right)\sin kx + \frac{kv}{\gamma^2+k^2v^2}e^{\gamma t}\cos kx\right].$$

It is seen that the two cases differ only in the initial out-of-phase contribution $f_1^s(t = 0)$, and given that $f_1^s(t = 0)$ is localised at the wave and that $\gamma(\partial f_0/\partial v)_w > 0$, these contributions differ only in sign.

**Energy Balance**

In the present formulation, calculation of the energy balance is straightforward.

The energy exchange is between the particles in the out-of-phase component of the distribution function $f_1$ and the electric field. The power transfer per unit volume, P, is

- 9 -

$$P = \int_1^s f \, ev \, E \, dv.$$

In the *frame of the wave*, $v_f = 0$, the contribution from particles with velocities close to the wave velocity is zero. So the only contribution is that from the basic plasma, and using eqn 5 with $\gamma \ll kv_w$

$$P = \int \frac{Ee}{m} \frac{\gamma}{k^2 v^2} \frac{\partial f_0}{\partial v} \cdot ev E dv.$$

Integrating by parts

$$P = \frac{E^2 e^2}{m} \frac{\gamma}{k^2} \int \frac{1}{v^2} f_0 \, dv$$

and putting $v = -v_w$,

$$P = \frac{E^2 e^2}{m} \frac{\gamma n}{k^2 v_w^2}.$$

Recalling that $E \propto e^{\gamma t}$, putting $k^2 v_w^2 = \omega_p^2$ and noting that $ne^2 / mk^2 v_w^2 = \varepsilon_0$, the energy balance equation becomes

$$P = \frac{d}{dt} \frac{\varepsilon_0 E^2}{2} \quad . \tag{13}$$

Thus, in the wave frame damping of the electrical energy is solely due to energy transfer to the particles in the main thermal distribution.

However, in the *frame of the plasma*, $v_f = -v_w$, there are two additional contributions, one from the basic plasma and the other from the particles around the wave speed. Using the sine component of the basic plasma density perturbation given by eqn 6

$$n_s^b = -\frac{2Een\gamma}{mk^2 v_w^3}$$

$$= -\frac{2\varepsilon_0 \gamma E}{ev_w}$$

the change in the power transferred by the basic plasma contribution is

$$\Delta P_b = -\frac{2\varepsilon_0 \gamma E}{ev_w} \cdot ev_w E$$

$$= -2\frac{d}{dt}\frac{\varepsilon_0 E^2}{2}.$$

So adding $\Delta P_b$ to the P given in eqn 13, $P_b$ in this frame is

$$P_b = -\frac{d}{dt}\frac{\varepsilon_0 E^2}{2}.$$

Since the out-of-phase density contribution of the particles close to the wave speed, $n_s^w$, is equal to $-n_s^b$, their power contribution, which was zero in the wave frame, is now

$$P_w = 2\frac{d}{dt}\frac{\varepsilon_0 E^2}{2}.$$

Summing the two contributions, the total power is

$$P = P_b + P_w$$

$$= \frac{d}{dt}\frac{\varepsilon_0 E^2}{2}$$

as before.

**Discussion**

In Landau's treatment, an arbitrary initial perturbation, $f_1(x,v,t = 0)$, of the distribution function is taken and the time development of the electric field, $E(x,t)$, is calculated using a Fourier-Laplace transform. Each component of $E(x,t)$ has a time dependence $e^{-i\omega t}$ with an eigenvalue $\omega(k)$. "Landau damping" is associated with an eigenvalue for which the real part of $\omega$ is such that $\omega_r/k \gg v_T$, the thermal velocity, and the imaginary part gives damping, or growth, proportional to $(\partial f_0/\partial v)_w$.

In Landau's derivation of the damping rate there is no involvement of the specific associated perturbed distribution function, and little attention has been paid to it. However, using the solution for E(x,t) the distribution function $f_1$(x,v,t) can be calculated retrospectively, thus identifying the specific initial distribution function required to give the Landau damping mode.

The present paper proceeds in a different way by solving the Vlasov equation directly to obtain $f_1$(k,v,t) and substituting this solution into Poisson's equation to obtain the dispersion relation giving the damping or growth rate.

For $(\partial f_0/\partial v)_w > 0$ the oscillation is unstable, and in this case the calculation proceeds straightforwardly. But in the damped case with $(\partial f_0/\partial v)_w < 0$, calculation of the $f_1$(k,v,t) requires the imposition of continuity with the $(\partial f_0/\partial v)_w > 0$ case, as in Landau's calculation.

The focus on the distribution function the present calculation makes explicit the underlying mechanisms of growth and damping. The growth associated with $(\partial f_0/\partial v)_w > 0$ arises from the charge produced by electrons with velocities sufficiently close to the wave velocity that they do not sample all phases of the wave during a growth time. The damping associated with $\partial f_0/\partial v)_w < 0$ is due to phase mixing of particles with velocities close to the wave velocity.

**Acknowledgment:** I am very grateful to Jeff Freidberg, Jack Connor and Jim Hastie for their interest, advice and encouragement.

**Reference:** 1. Landau, L.D. *Journal of Physics (USSR)* **10,** 25 (1946).